\documentclass[superscriptaddress,amsmath,amssymb,nofootinbib,longbibliography]{revtex4-1}
\pdfoutput=1

\usepackage[T1]{fontenc} 
\usepackage{amssymb,amsmath,graphicx,amscd,xcolor,amsthm}
\usepackage{braket}
\usepackage{verbatim} 
\usepackage[colorlinks,citecolor=blue,urlcolor=blue]{hyperref}
\usepackage[caption=false]{subfig}

\usepackage{braket}

\newcommand{\de}{\mr{d}}
\newcommand{\x}{\mathbf{x}}
\newcommand{\y}{\mathbf{y}}
\newcommand{\z}{\mathbf{z}}

\newcommand{\p}{\mathbf{p}}
\newcommand{\vv}{\mathbf{v}}
\newcommand{\E}{\mathbf{E}}
\newcommand{\vL}{\mathbf{L}}
\newcommand{\F}{\mathbf{F}}

\newcommand{\mr}[1]{\mathrm{#1}}
\newcommand{\mat}[1]{\mbox{\boldmath{$#1$}}}

\makeatletter
\@booleanfalse\titlepage@sw
\makeatother

\begin{document}

\title{Anisotropic motion of an electric dipole in a photon gas near a flat conducting boundary}

\author{G. H. S. \surname{Camargo}}
\email{guilhermehenrique@unifei.edu.br}
\affiliation{Instituto de Ci\^encias Exatas, Universidade Federal de Juiz de Fora, Juiz de Fora, Minas Gerais 36036-330,  Brazil}
\affiliation{Instituto de F\'{\i}sica e Qu\'{\i}mica, Universidade Federal de Itajub\'a, Itajub\'a, Minas Gerais 37500-903, Brazil}

\author{V. A. \surname{De Lorenci}}
\email{delorenci@unifei.edu.br}
\affiliation{Instituto de F\'{\i}sica e Qu\'{\i}mica, Universidade Federal de Itajub\'a, Itajub\'a, Minas Gerais 37500-903, Brazil}
\affiliation{{${\cal G}\mathbb{R}\varepsilon\mathbb{C}{\cal
      O}$}---Institut d'Astrophysique de Paris, CNRS \& Sorbonne
  Universit\'e, UMR 7095 98 bis Boulevard Arago, 75014 Paris, France}
\author{A. L.  \surname{Ferreira Junior}}
\email{alexsandre.ferreira@edu.ufes.br}
\affiliation{PPGCosmo, CCE - Universidade Federal do Esp\'{\i}rito Santo, Vit\'{o}ria, Esp\'{\i}rito Santo, 29075-910, Brazil.}
\author{C. C. H. \surname{Ribeiro}}
\email{caiocesarribeiro@ifsc.usp.br}
\affiliation{Centro Internacional de F\'isica, Instituto de F\'isica, Universidade de Bras\'ilia, 70910-900, Bras\'ilia, DF, Brazil}

\begin{abstract}
The quantum Brownian motion of a single neutral particle with nonzero electric dipole moment placed in a photon gas at fixed temperature and close to a conducting wall is here examined. The interaction of the particle with the photon field leads to quantum dispersions of its linear and angular momenta, whose magnitudes depend on the temperature, distance to the wall, and also on the dipole moment characteristics. It is shown that for typical experimental parameters the amount of energy held by the dipole rotation is expressively larger than the one related to the center of mass translation. Furthermore, the particle kinetic energy in presence of a thermal bath can decrease if the wall is added to the system, representing a novel quantum cooling effect where the work done by the quantum vacuum extracts energy from the particle. Finally, possible observable consequences are discussed.
\end{abstract}

\maketitle
\flushbottom

\section{Introduction}
Decoherence represents a natural mechanism for the emergence of a classical notion of observables from quantum systems and can occur, for instance, in systems interacting with thermal environments \cite{schlosshauer2019}. In the latter, fluctuations induced by the environment suppress quantum interference effects, turning the quantum system into a statistical mixture on a time scale smaller than the one the system takes to reach equilibrium (see Ref.~\cite{unruh1989}, and also \cite{Vanzella2015} for an example in semiclassical gravity). Contributing to its importance, it should be noted that this type of interaction may be prevalent in our universe, as probably all naturally occurring quantum systems experience decoherence, i.e., every system is immersed in a thermal bath of long wavelength gravitons \cite{delorenci2015}.

It is then clear that if one is interested in quantum-based technology, one must have full control of the thermal fluctuations of quantum fields. Examples of fundamental phenomena such as the Hawking and Unruh effects \cite{Hawking74,Unruh1976} show that a deep understanding of all existing thermal contributions is necessary in order to disentangle the former in a measured signal, which remains an essentially unfeasible task. Within the program of probing (weak) semiclassical gravity phenomena, theoretical models and alternative measuring procedures are constantly being explored by the scientific community. We cite for instance the models for probing the Unruh effect \cite{Vanzella2017,Vanzella2019}, and the recent measurement of Hawking-like radiation in the context of analogue gravity \cite{Steinhauer2019}, which is an important milestone. Moreover, of particular importance for the present work is the notion of subvacuum phenomena \cite{delorenci2019,Jaffe,Hsiang,Fordsub}, for which classically positive quantities assume negative values after renormalization. It was recently shown \cite{camargo2021} that temperature can enhance subvacuum effects in some systems of boundary physics. Yet, the possibility of detection requires the thermal fluctuations to be well-known and distinguished from the boundary contributions.

Examining the behaviour of  systems under influence of thermal fluctuations can also lead to conclusions about the coupling between gravity and physical fields. In fact, first principles analysis shows that the interplay between vacuum and thermal local averages for scalar radiation near a reflecting wall implies in a natural restriction on the possible values the curvature coupling parameter $\xi$ can take \cite{delorenci2015b} for the system to be thermodynamically stable. Particularly, it was found that in more than three spacetime dimensions such a range contains the conformal coupling, but it does not contain the minimal coupling. 

In this paper, we study quantum fluctuations of the radiation field through a test neutral particle with nonzero dipolar moment, which models one of the most usual couplings between radiation and ordinary matter. We work in the stochastic regime, in which the decoherence has already played its part and the particle is treated semiclassically, i.e., only the electromagnetic field is quantized. Such models are well suited to investigate the construction of a bottom-up thermodynamics \cite{hu2018,hu2019}.

Due to the field-matter interaction, aspects of the background field can be tested through the motion of the test particle, and this induced motion resembles the modified vacuum induced motion studied in Refs. \cite{camargo2021,bessa2009,ford2004,delorenci2014,delorenci2016,Camargo2018,delorenci2019b,yu2006,jt2009}. In these works, a point-like charged particle was shown to perform a sort of random walk due to a transition between states of the fields. Furthermore, for the particular case of Ref. \cite{delorenci2016}, the charged particle entered a region in which the electromagnetic vacuum state was modified by the presence of a plane perfectly conductor. That models a transition between a homogeneous isotropic background to one in which both symmetries are broken by the presence of the boundary, which is thus reflected by an anisotropy in the particle's random walk. The behaviour of a charged particle in a finite temperature regime was discussed in Refs. \cite{yu2006,jt2009,delorenci2019b,camargo2021}.

In the present work the dipolar particle experiences vacuum fluctuations of a thermal bath near a perfectly conducting plane. Within this model, by adjusting the temperature of the radiation and the particle's distance to the wall, the effects of the isotropic thermal bath and of the wall over the particle's motion can be studied separately. Because the dipole moment of the particle breaks the isotropy of the system, even for the case of a thermal bath only, an anisotropic motion of the particle is expected to occur.

Natural units $\hbar =1$, $c =1=\varepsilon_0$, and $\kappa_{B}=1$ are used throughout this manuscript, except where estimates of the effects are presented. 

\section{Preliminary aspects}
\label{secII}
The main interest here is the motion of a non-relativistic neutral particle of mass $m$ and electric dipole moment $\p$. For definiteness we neglect the internal structure of the dipole, and consider the case where the particle is placed in the presence of an external electric field $\E_{\tt ext}(\x,t)$ such that $U(\x,t)=-\p\cdot\E_{\tt ext}(\x,t)$ is its potential energy. 

Hereafter, the electric field $\E_{\tt ext}$ is taken to be composed of a classical plus a fluctuating field $\E_{\rm c}+\E$, where the classical field $\E_{\rm c}$ is set as to keep the dipole (optically) trapped at a given trajectory and direction \cite{Ashkin1986,Bustamante}, i.e., we have $\nabla (\p\cdot\E_{\rm c})=0$ and $\p\times\E_{\rm c}=0$ at the particle's position throughout the system evolution. Accordingly, the force and torque experienced by the particle are  given by  $\F(\x,t) = - \mat{\nabla} U(\x,t)$ and
 $\mathbf{T}(\x,t) = \p\times\E(\x,t)$, respectively, where $\mathbf{T}$ is defined with respect to its center of mass at $\x$. Therefore, the particle motion is described by the equations
\begin{align}
m\frac{\de\vv}{\de t}&=\nabla (\p\cdot\E),\label{newton1}\\
\frac{\de\vL}{\de t}&= \p\times\E\label{newton2},
\end{align}
where the particle angular momentum $\vL$ is also defined with respect to the dipole center of mass. Furthermore, the trapped dipole is assumed to be near a perfectly conducting plane in thermal equilibrium, where the classical field $\E_{\rm c}$ is adjusted accordingly in order to counteract classical backreaction effects coming from the dipole interaction with the mirror. Also, we assume here the (test) particle coupling to the electromagnetic field is sufficiently weak such that the thermal equilibrium is unperturbed, and the particle motion occurs solely due to fluctuating field $\E$. 
In what follows, we work in a regime where variations in the position of the dipole center of mass and dipole orientation caused by the interaction with $\E$ can be neglected. This last assumption sets a time scale for the applicability of our model, and it is justified in view of the smallness of the quantum effects \cite{delorenci2016}. We return to this point when we discuss estimates in section \ref{estimates}. Thus, it follows from Eq.~\eqref{newton1} that the particle velocity is (in Cartesian components)
\begin{equation}
v_i(\tau)=\sum_{j=1}^3\frac{p_j}{m}\int_0^{\tau}\de t\,\partial_i E_j(\x,t),
\label{vi}
\end{equation}
where we took $p_j$ out of the integral following the time-independence of $\p$. 

\begin{figure}[tbp]
\center
\includegraphics[width=0.45\textwidth]{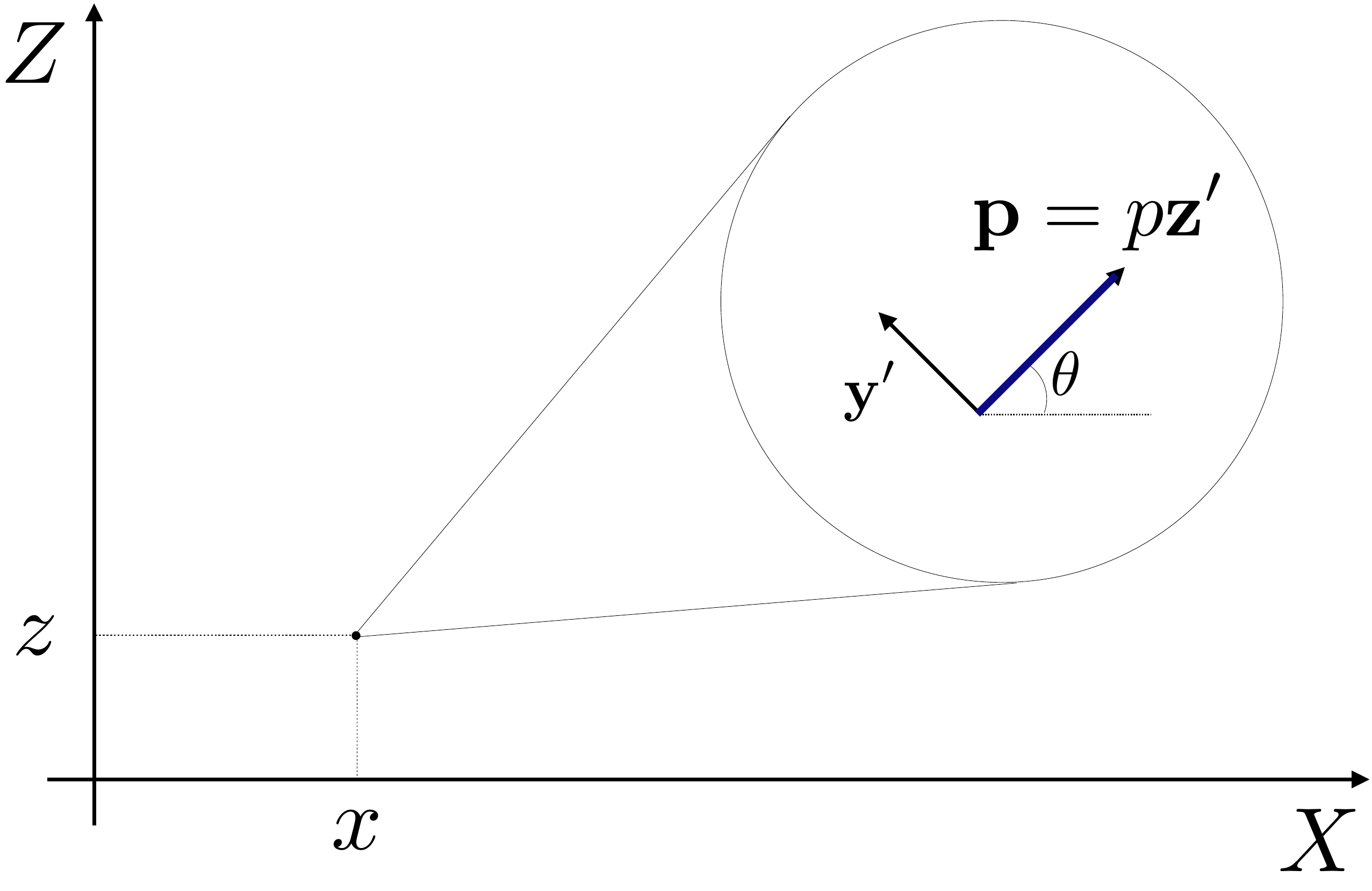}
\caption{System arrangement, showing the choice of coordinates we use to describe the dipole motion. In particular, we set $\p=p_x\hat{\x}+p_z\hat{\z}=p(\cos\theta\hat{\x}+\sin\theta\hat{\z})$. The $\hat{\y}$ vector is entering the plane of the page, and we recall that the dipole is (classically) forced to maintain this configuration.}
\label{figangle}
\end{figure}

Analogously, from Eq.~\eqref{newton2} we can find an equation satisfied by the components $\omega_i$ of the particle angular velocity by exploring the general relation $L_i=I_{ij}\omega_j$, where $I_{ij}$ are the components of the inertia tensor \cite{landaumec}. We take the mirror to coincide with the plane $z=0$, and the dipole placed at a distance $z$ from this plane as shown in Figure~\ref{figangle}.
We also assume that the dipole is a symmetric top whose symmetry axis is parallel to $\p$. For the arrangement shown in Figure~\ref{figangle}, we can thus fix a coordinate system to the dipole that instantaneously coincide with $\{\hat{\x}',\hat{\y}',\hat{\z}'\}$, where $\hat{\x}'=\hat{\y}, \hat{\y}'=\hat{\p}\times\hat{\y}, \hat{\z}'=\hat{\p}$, $\hat{\p}=\p/p$. Accordingly, $I_{ij}$ expressed in terms of $\{\hat{\x}',\hat{\y}',\hat{\z}'\}$ assumes the diagonal form $I_{ij}=\mbox{diag}(I_\bot,I_\bot,I_\parallel)$, and thus Eq.~\eqref{newton2} reduce to the Euler equations \cite{landaumec}
\begin{align}
&I_{\bot}\frac{\de\omega_{x'}}{\de t}-(I_\bot-I_\parallel)\omega_{y'}\omega_{z'}=p_zE_x-p_xE_z,\label{angx}\\
&I_{\bot}\frac{\de\omega_{y'}}{\de t}-(I_\parallel-I_\bot)\omega_{x'}\omega_{z'}=pE_y,\label{angy}\\
&I_{\parallel}\frac{\de\omega_{z'}}{\de t}=0.
\end{align}
Therefore, the symmetry of the dipole along its axis and the particular form of the torque $\p\times\E$ results in $\omega_{z'}$ constant, from which we take $\omega_{z'}=0$. With this initial condition added, it thus follows from Eqs.~\eqref{angx} and \eqref{angy} that
\begin{align}
&\omega_{x'}(\tau)=\frac{p}{I}\int_{0}^{\tau}\de t\left(E_x\sin\theta-E_z\cos\theta\right),\label{velang1}\\
&\omega_{y'}(\tau)=\frac{p}{I}\int_{0}^{\tau}\de tE_y\label{velang2},
\end{align}
where we set $I:=I_{\bot}$. We observe that Eqs.~\eqref{velang1} and \eqref{velang2} assume simplified forms {\it only} when $\omega_{z'}=0$. We observe that when $\theta=\pi/2$, i.e., the dipole moment is perpendicular to the mirror, the symmetry of the problem  shows that both $\omega_{x'}$ and $\omega_{y'}$ represent the same type of rotation with respect to the plane $z=0$, whereas when $\theta=0$, $\omega_{y'}$ still has the same interpretation but $\omega_{x'}$ then corresponds to a distinct rotation of the dipole moment parallel to the plane $z=0$.

Finally, Eqs.~\eqref{vi}, \eqref{velang1}, and \eqref{velang2} suppose the effect of $\E$ on the particle is suddenly turned on and off at $t=0$ and $\tau$, respectively. As pointed out in \cite{delorenci2019b}, this assumption leads to velocity dispersion divergences for the case of an electric monopole, and as we shall see the same phenomenon is present also for the dipole.

In the following we take the field $\E$ to be quantized in a quantum state modeling the electromagnetic fluctuations at thermal equilibrium near the mirror at $z=0$.
Accordingly, the particle velocities $v_i$ and $\omega_i$ become operators through their dependence on $\E$, thus giving rise to a semiclassical description of its motion, i.e., we are neglecting effects coming from the particle's wave function.
As in any observable in a quantum system, measurement predictions are given through expectation values of these operators. Moreover, as we are dealing with a field in thermal equilibrium, the expectation value of an observable $O$ is given by $\braket{O}=\mbox{tr}[\rho\,O]$. The density operator $\rho= (1/Z)\textrm{exp}\left(-\beta H\right)$ describes the grand-canonical ensemble (zero chemical potential) with temperature $T\doteq 1/\beta$, where $Z=\sum_i\textrm{exp}\left[-\beta \mathcal{E}_i\right]$ denotes the partition function, and $\{\mathcal{E}_i\}$ the set of eigenvalues of the Hamiltonian $H$ \cite{Davies1982}.

Notice that although the radiation is in thermal equilibrium, the particle, assumed to depart from a regime of zero kinetic energy, is not.
Yet, the average force and torque on the particle is zero because of the system quantum state: $\braket{E_i}=0$. It also follows that $\braket{v_i}=0$ and $\braket{\omega_i}=0$. Therefore, the squared mean deviations are simply $\braket{(\Delta v_i)^2}=\braket{v_i{}^2}$ and $\braket{(\Delta \omega_i)^2}=\braket{\omega_i{}^2}$, and these quantities describe how measurements of the corresponding observables are distributed around their zero average. Furthermore, we stress that our working assumption used to derive Eqs.~\eqref{vi}, \eqref{velang1}, and \eqref{velang2} prevents the formal limit $\tau\rightarrow\infty$ from being studied in our analysis, the reason for it being the particle (quantum) Brownian motion. Indeed, velocity fluctuations correspond to position dispersions in a similar fashion as occurs in the classical Brownian motion \cite{li2010}, and because Eqs.~\eqref{vi}, \eqref{velang1}, and \eqref{velang2} are valid as long as the particle remains motionless with respect to the laboratory frame, the measuring time $\tau$ is bounded. We discuss limits on $\tau$ for the validity of our results in section \ref{estimates}.
Therefore, taking the expectation value of the square of $v_i$, given by Eq.~\eqref{vi}, we obtain,
\begin{equation}
    \braket{{v_i}^2}=\sum_{j=1}^3\frac{p_j^2}{m^2}\lim_{\x'\to\x}\int_0^\tau\int_0^\tau\de t\de t'\,\partial_i\partial_{i'} \braket{E_j(\x,t)E_j(\x',t')},
    \label{corv}
\end{equation}
and similarly with Eqs.~\eqref{velang1} and \eqref{velang2},
\begin{align}
    \braket{{\omega_{x'}}^2}=&\frac{p^2}{I^2}\int_0^\tau\int_0^\tau\de t\de t'\, \left[\braket{E_x(\x,t)E_x(\x,t')}\sin^2\theta+\braket{E_z(\x,t)E_z(\x,t')}\cos^2\theta\right],\label{corw1}\\
    \
    \braket{{\omega_{y'}}^2}=&\frac{p^2}{I^2}\int_0^\tau\int_0^\tau\de t\de t'\, \braket{E_y(\x,t)E_y(\x,t')}.\label{corw2}
\end{align}

Finally, when calculating the dispersions, we need to know the quantum correlations for the electromagnetic field $\langle E_i(t,\x)E_i(t',\x')\rangle$, $i=x,y,z$, which can be read from \cite{Brown}. In particular, we identify three separate contributions for the dispersion as $\langle E_x(t,\x)E_x(t',\x')\rangle=\langle E_x(t,\x)E_x(t',\x')\rangle_{\rm th}+\langle E_x(t,\x)E_x(t',\x')\rangle_{\rm vc}+\langle E_x(t,\x)E_x(t',\x')\rangle_{\rm mx}$ and similarly for $y$ and $z$ components as shown in the Appendix \ref{appA}. Hereafter, the subindices ``$\rm th$'', ``$\rm vc$'', and ``$\rm mx$'' stand for thermal, vacuum, and mixed, respectively. These contributions are such that when $T\rightarrow0$, only $\langle E_x(t,\x)E_x(t',\x')\rangle_{\rm vc}$ remains nonzero, representing thus the effect of the mirror, whereas if $z\rightarrow\infty$, i.e., if the particle is far away from the mirror, only $\langle E_x(t,\x)E_x(t',\x')\rangle_{\rm th}$ contributes, identifying the effect of the isotropic thermal bath. Finally, $\langle E_x(t,\x)E_x(t',\x')\rangle_{\rm mx}$ corresponds to a mixed contribution from mirror plus thermal bath. Therefore, each velocity dispersion also admits the same splitting, which we investigate separately in what follows.

\section{Velocity dispersions}

\subsection{Thermal fluctuations}
\label{thermalfluc}
\begin{figure}[tbp]
\center
\includegraphics[width=0.5\textwidth]{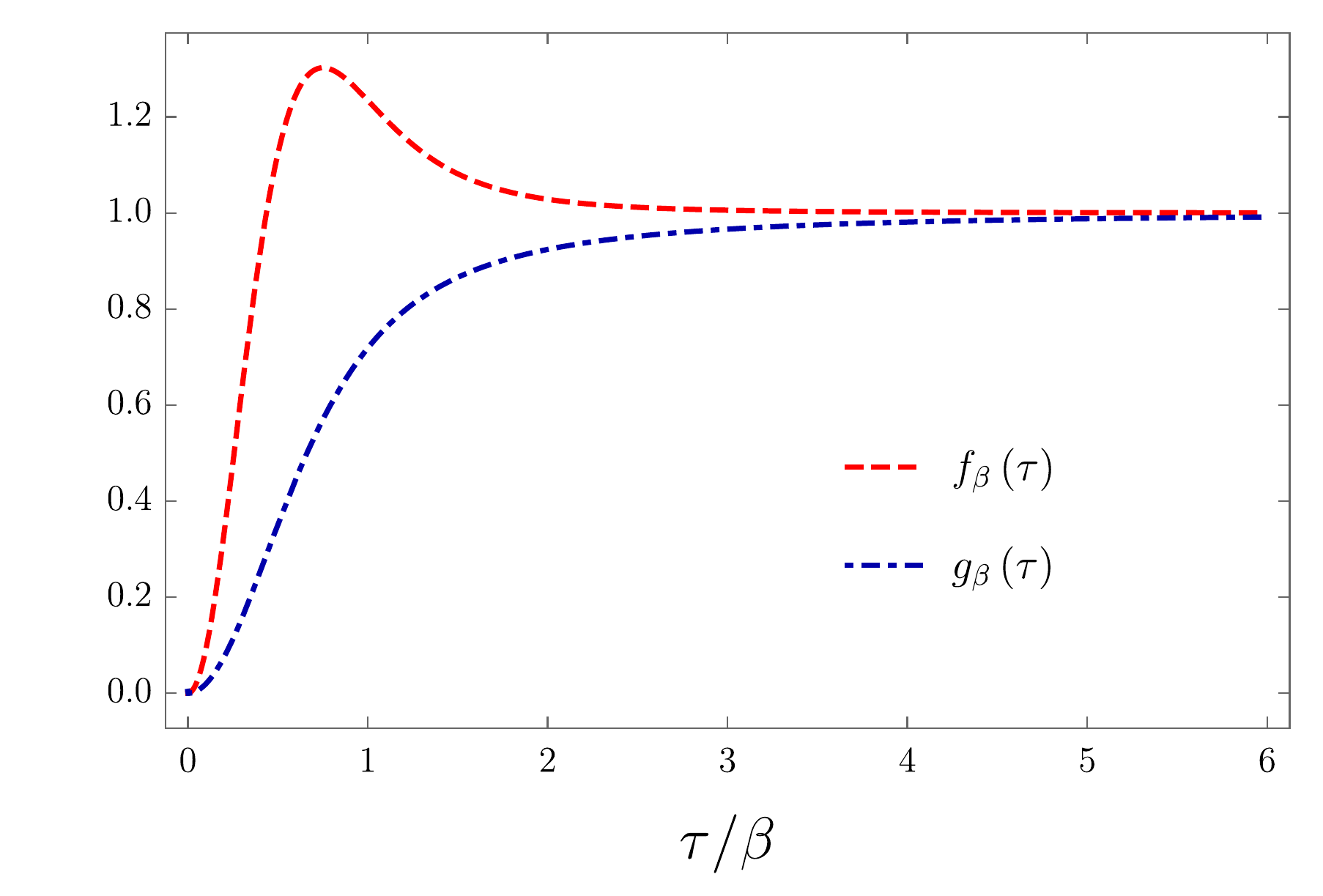}
\caption{Behaviour of the auxiliary functions $f_\beta(\tau)$ and $g_\beta(\tau)$ as function of time. They vanish for $\tau/\beta\rightarrow0$ and tend to $1$ as $\tau/\beta\rightarrow\infty$.}
\label{fig2}
\end{figure}

We start by considering the effect of the thermal bath alone, i.e., the case in which the dipole is far from the wall at $z=0$. We present in the Appendix \ref{appA} all the relevant formulas to calculate the dispersions. Due to the fact that the dipole moment of the particle breaks isotropy, the quantum dispersion of the particle velocity ${\mathbf v}$ is anisotropic. Specifically, the dispersion of the velocity component perpendicular to the dipole axis $\braket{v_{x'}{}^2}_{\rm th}$ is twice the dispersion along its axis $\braket{v_{z'}{}^2}_{\rm th}$, whereas for the angular velocity we find that $\braket{\omega_{x'}^2}_{\rm th}=\braket{\omega_{y'}^2}_{\rm th}$. Hence, the mean value of the velocities squared are $\braket{\vv^2}_{\rm th} =  5\braket{v_{z'}{}^2}_{\rm th}$ and $\langle{\boldsymbol{\omega}}^2\rangle_{\rm th}=2\braket{\omega_{x'}^2}_{\rm th}$, where
\begin{align}
\braket{\vv{}^2}_{\rm th}&=\frac{2\pi^2p^2}{45 m^2 \beta^4} f_{\beta}(\tau),
\label{v}
\\
\langle{\boldsymbol{\omega}}^2\rangle_{\rm th}&=\frac{2p^2}{9 I^2 \beta^2}\, g_{\beta}(\tau),
\label{omega}
\end{align}
and we have defined the dimensionless functions of $\tau$,
\begin{align}
f_\beta(\tau)&=1+\frac{45\beta^4}{\pi^4\tau^4} -15\left[2+\cosh\left(\frac{2\pi\tau}{\beta}\right)\right]{\rm csch}^4\!\left(\frac{\pi\tau}{\beta}\right),
\nonumber
\\
 g_\beta(\tau) &=1-\frac{3\beta^2}{\pi^2\tau^2}+3  {\rm csch}^2\!\left(\frac{\pi\tau}{\beta}\right).
 \nonumber
\end{align}
The auxiliary functions $f_\beta(\tau)$ and $g_\beta(\tau)$ are depicted in Figure~\ref{fig2}.

In particular, no $\tau$-dependent divergence is observed in $\braket{\vv{}^2}_{\rm th}, \langle{\boldsymbol{\omega}}^2\rangle_{\rm th}$, which is expected to occur only when the contributions from the mirror are added.
Notice that both functions $f_\beta(\tau)$ and $g_\beta(\tau)$ vanish when $\tau \to 0$ and go to 1 when $\tau \to \infty$. Thus, the factors multiplying these functions in Eqs.~(\ref{v}) and (\ref{omega}) correspond to the asymptotic values of the velocity dispersions.
As we are assuming the sudden regime approximation, the calculated dispersions correspond to a particle which is placed initially (at $t=0s$) at rest in the bulk of a blackbody cavity at a temperature $T$, where a Brownian motion induced by the radiation field begins according to Eqs.~\eqref{v} and \eqref{omega}. We observe that both translational and rotational velocity dispersions tend to asymptotic values as $\tau\rightarrow\infty$, suggesting that the system enters a ``quantum ballistic'' regime in analogy with the classical Brownian motion \cite{li2010}. However, we recall that our results are reliable as long as the dipole moment $\p$ remains time-independent, which is certainly not the case in the limit $\tau\rightarrow\infty$, where a constant nonzero angular velocity is reached. We return to this discussion in section \ref{estimates}.

\subsection{Zero temperature fluctuations}
\label{zerofluc}

\subsubsection{Translational velocity dispersions}

We now consider the opposite regime where only the mirror induces velocity dispersion, i.e., we take $T=0$. For this case we find that
\begin{align}
&\langle v_x^2\rangle_{\rm vc}=\frac{p^2}{8\pi^2m^2z^4}\left[(\delta v_x)^2_{\rm vc,\parallel}
\cos^2\theta+(\delta v_x)^2_{\rm vc,\bot}
\sin^2\theta\right],\label{eqqq1}\\
&\langle v_y^2\rangle_{\rm vc}=\frac{p^2}{8\pi^2m^2z^4}\left[(\delta v_y)^2_{\rm vc,\parallel}
\cos^2\theta+(\delta v_x)^2_{\rm vc,\bot}
\sin^2\theta\right],\\
&\langle v_z^2\rangle_{\rm vc}=\frac{p^2}{8\pi^2m^2z^4}\left[(\delta v_z)^2_{\rm vc,\parallel}
\cos^2\theta+(\delta v_z)^2_{\rm vc,\bot}
\sin^2\theta\right],
\end{align}
where
\begin{subequations}
\label{gene}
\begin{align}
&(\delta v_x)^2_{\rm vc,\parallel}=\frac{1}{4}\left[\eta^2\frac{7-5\eta^2}{(\eta^2-1)^2}+\frac{9\eta}{2}\ln\left|\frac{\eta+1}{\eta-1}\right|\right],\\
&(\delta v_x)^2_{\rm vc,\bot}=-\frac{\eta^2}{\eta^2-1}+\frac{3\eta}{2}\ln\left|\frac{\eta+1}{\eta-1}\right|,\\
&(\delta v_y)^2_{\rm vc,\parallel}=\frac{1}{4}\left[\eta^2\frac{5-3\eta^2}{(\eta^2-1)^2}+\frac{3\eta}{2}\ln\left|\frac{\eta+1}{\eta-1}\right|\right],\\
&(\delta v_z)^2_{\rm vc,\parallel}=\eta^2\frac{7-8\eta^2+3\eta^4}{(1-\eta^2)^3}+\frac{3\eta}{2}\ln\left|\frac{\eta+1}{\eta-1}\right|,\\
&(\delta v_z)^2_{\rm vc,\bot}=\eta^2\frac{4-3\eta^2}{(\eta^2-1)^2}+3\eta\ln\left|\frac{\eta+1}{\eta-1}\right|,
\end{align}
\end{subequations}
and $\eta=\tau/(2z)$. Apart from the factor $p^2/(8\pi^2m^2z^4)$, the function $(\delta v_x)^2_{\rm vc,\parallel}$ in Eq.~\eqref{eqqq1} represents the induced velocity dispersion in the $x$ direction for the dipole oriented along the $x$ axis (cf. Figure~\ref{figangle} for $\theta=0$), and similarly for $(\delta v_x)^2_{\rm vc,\bot}$, $(\delta v_y)^2_{\rm vc,\parallel}$, $(\delta v_z)^2_{\rm vc,\parallel}$, $(\delta v_z)^2_{\rm vc,\bot}$.
Equations \eqref{gene} extend the results presented in \cite{ford2004} for particles with nonzero dipolar moments. We observe that, in agreement with the monopole analysis, the velocity dispersions diverge at $\tau=2z$ ($\eta=1$), the time taken for a light signal to round trip between the particle and the mirror. In general, the origin of this divergence comes from the assumptions that the fluctuating field $\E$ is zero immediately before $t=0$ and after $t=\tau$, the radiation reaction from the particle acceleration can be neglected, and the mirror is perfect \cite{Ribeiro2023}. For instance, a smooth transition between field states can be modeled with switching functions, capable of regularising the dispersions for the case of electric monopoles \cite{delorenci2019b}.

\begin{figure}[tbp]
\center
\includegraphics[width=0.5\textwidth]{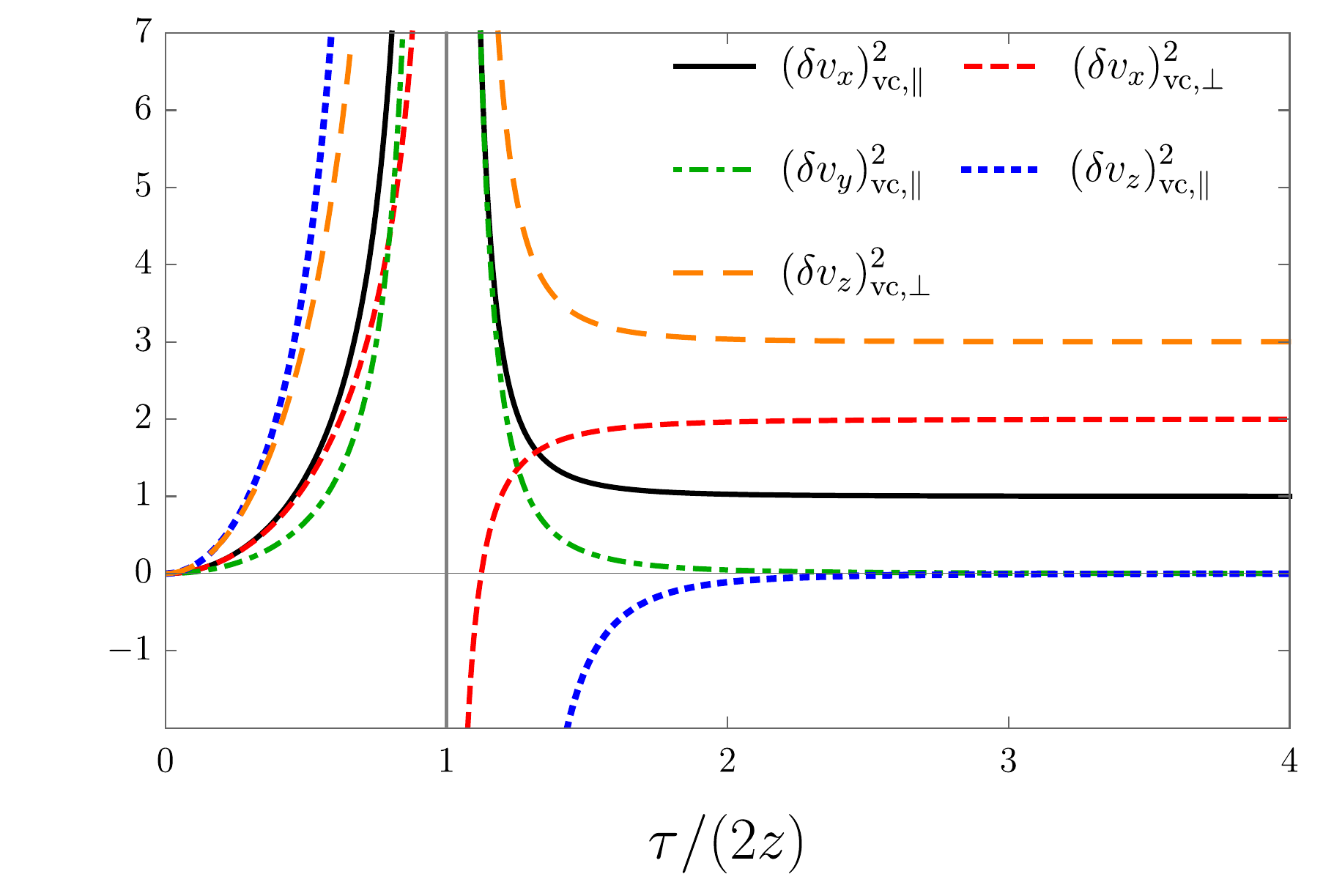}
\caption{Translational velocity dispersions as function of $\tau/(2z)$ (up to a factor) at $T=0$. We observe the characteristic divergence at $\tau=2z$, which is also present for electric monopoles, and is more pronounced in the case of dipolar particles. Moreover, the dispersions quickly enter a stationary regime after $\tau=2z$. These divergences are naturally regularised when a smooth switching is considered \cite{delorenci2019b}.}
\label{figC}
\end{figure}

Figure \ref{figC} depicts our findings for the functions defined in Eqs.~\eqref{gene}.
We observe that the dispersions quickly stabilise at their asymptotic values $(\delta v_x)^2_{\rm vc,\parallel}\rightarrow1$, $(\delta v_x)^2_{\rm vc,\bot}\rightarrow2$, $(\delta v_y)^2_{\rm vc,\parallel}\rightarrow0$, $(\delta v_z)^2_{\rm vc,\parallel}\rightarrow0$, and $(\delta v_z)^2_{\rm vc,\bot}\rightarrow3$ as $\eta\rightarrow\infty$. Furthermore, in the previous section we showed how the dipole moment is associated to anisotropic dispersions even for isotropic thermal baths, and naturally this behaviour is more pronounced in anisotropic vacuum states as the one modeled by the mirror, as revealed by Eqs.~\eqref{gene}. In particular, the divergence at $\eta=1$ is weaker for the velocity dispersion parallel to the mirror for the dipole oriented along the $z$ axis, and it is stronger in the opposite configuration: Perpendicular motion of a particle whose dipole moment is parallel to the mirror. Therefore, because dipolar particles are more sensitive to vacuum fluctuations than monopoles, our analysis offers a distinct route for probing field correlations which is within experimental reach.

\subsubsection{Angular velocity dispersion}

\begin{figure}[tbp]
\center
\includegraphics[width=0.5\textwidth]{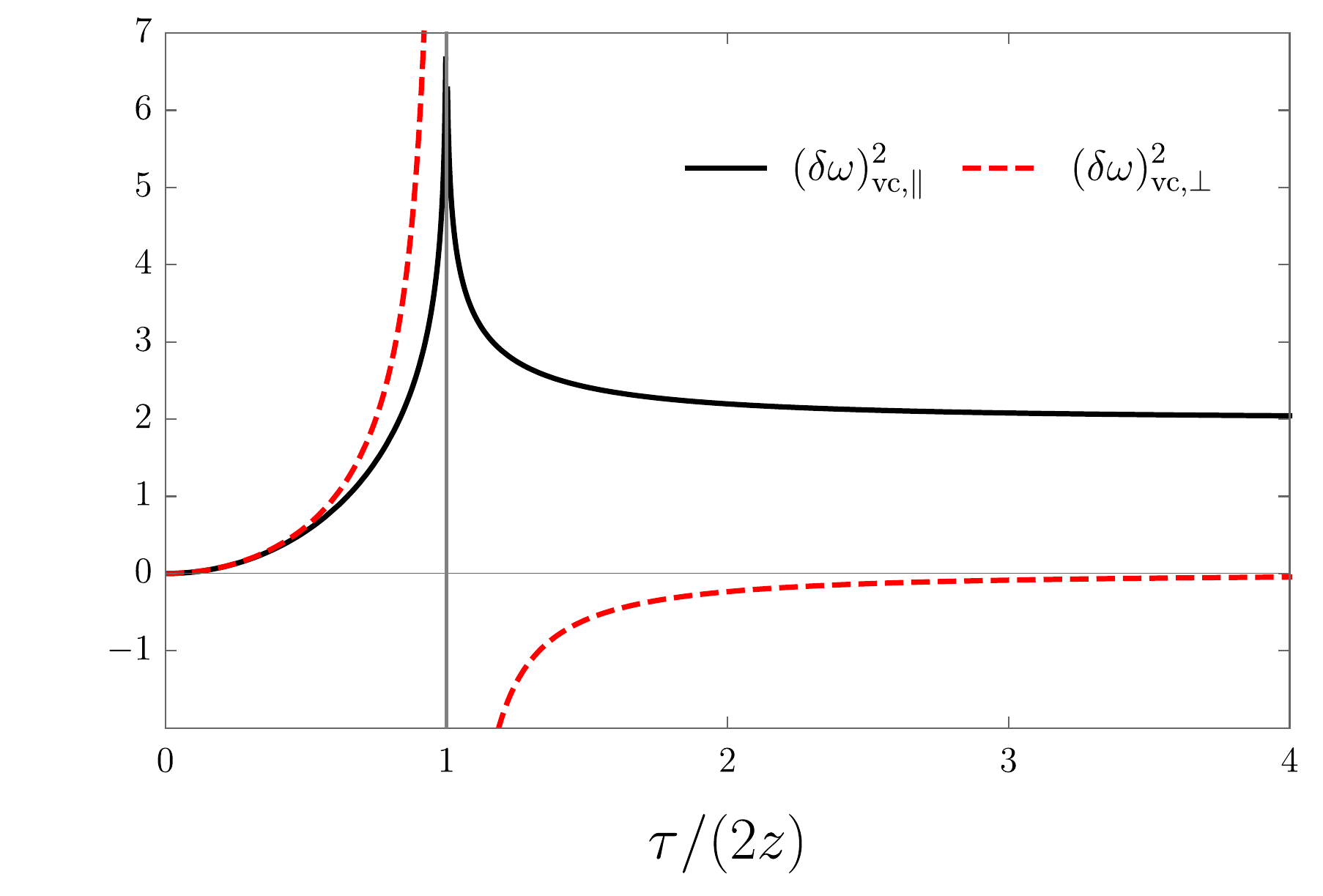}
\caption{Angular velocity dispersions (up to a factor) at $T=0$. The continuous curve represents the angular velocity of the dipole moment vector when it is aligned parallel to the wall ($\theta=0$), whereas the dashed curve corresponds to a rotation perpendicular to the wall ($\theta=\pi/2$) (cf. Figure~\ref{figangle}). Similarly to the translational velocity dispersions, we see that the dispersions diverge at $\tau=2z$.}
\label{figD}
\end{figure}

From the results in the Appendix \ref{appA}, we find that the angular velocity dispersions when $T=0$ assume the form
\begin{align}
&\langle\omega_{x'}^2\rangle_{\rm vc}=\frac{p^2}{8I^2\pi^2z^2}\left[(\delta\omega)^2_{\rm vc,\parallel}\cos^2\theta+(\delta\omega)^2_{\rm vc,\bot}\sin^2\theta\right],\\
&\langle\omega_{y'}^2\rangle_{\rm vc}=\frac{p^2}{8I^2\pi^2z^2}(\delta\omega)^2_{\rm vc,\bot},
\end{align}
where
\begin{align}
(\delta\omega)^2_{\rm vc,\parallel}&=\eta\ln\left|\frac{\eta+1}{\eta-1}\right|,\\
(\delta\omega)^2_{\rm vc,\bot}&=-\frac{\eta^2}{\eta^2-1}+\frac{\eta}{2}\ln\left|\frac{\eta+1}{\eta-1}\right|.
\end{align}
Apart from the factor $p^2/(8I^2\pi^2z^2)$, the function $(\delta\omega)^2_{\rm vc,\bot}$ corresponds to the dispersion of the angular velocity of a dipole oriented along the $z$ axis, i.e., perpendicular to the conductor at $z=0$, whereas $(\delta\omega)^2_{\rm vc,\parallel}$ is the dispersion of the angular velocity along the $y$ axis for the dipole moment oriented parallel to the plane ($\theta=0$). Both functions $(\delta\omega)^2_{\rm vc,\bot}$, $(\delta\omega)^2_{\rm vc,\bot}$ are readily seem to diverge at $\eta=1$, behave as $(\delta\omega)^2_{\rm vc,\parallel}\rightarrow 2$, $(\delta\omega)^2_{\rm vc,\bot}\rightarrow 0$ as $\eta\rightarrow\infty$, and are depicted in Figure~\ref{figD}.
We also note that the mathematical expressions for angular velocity dispersions coincide with the translational velocity dispersions for the analogous case of an electric monopole \cite{ford2004}, with $p$ playing the role of the electric charge and $I$ of the particle mass.

\subsection{Mixed fluctuations}

\subsubsection{Translational velocity dispersions}

The remaining contribution to the velocity dispersions exists {\it only} when both $T\neq0$ and the mirror is present. The contribution to the particle velocity dispersion due to the mixed fluctuations can be written as
\begin{align}
&\langle v_x^2\rangle_{\rm mx}=\frac{p^2}{16\pi^2m^2z^4}\left[(\delta v_x)^2_{\rm mx,\parallel}
\cos^2\theta+(\delta v_x)^2_{\rm mx,\bot}
\sin^2\theta\right],\\
&\langle v_y^2\rangle_{\rm mx}=\frac{p^2}{16\pi^2m^2z^4}\left[(\delta v_y)^2_{\rm mx,\parallel}
\cos^2\theta+(\delta v_x)^2_{\rm mx,\bot}
\sin^2\theta\right],\\
&\langle v_z^2\rangle_{\rm mx}=\frac{p^2}{16\pi^2m^2z^4}\left[(\delta v_z)^2_{\rm mx,\parallel}
\cos^2\theta+(\delta v_z)^2_{\rm mx,\bot}
\sin^2\theta\right],
\end{align}
where
\begin{subequations}
\label{velmixed}
\begin{align}
(\delta v_x)^2_{\rm mx,\parallel}=&\frac{\beta}{4z}\mbox{Im}\left[\left(9-9z\partial_z+4z^2\partial_z^2-z^3\partial^3_z\right)f\right],\\
(\delta v_x)^2_{\rm mx,\bot}=&\frac{\beta}{z}\mbox{Im}\left[\left(3-3z\partial_z+z^2\partial_z^2\right)f\right],\\
(\delta v_y)^2_{\rm mx,\parallel}=&\frac{\beta}{4z}\mbox{Im}\left[\left(3-3z\partial_z+2z^2\partial_z^2-z^3\partial^3_z\right)f\right],\\
(\delta v_z)^2_{\rm mx,\parallel}=&\frac{\beta}{z}\mbox{Im}\bigg[\bigg(3-3z\partial_z+\frac{7}{4}z^2\partial_z^2-\frac{3}{4}z^3\partial^3_z+\frac{1}{4}z^4\partial^4_z\bigg)f\bigg],\\
(\delta v_z)^2_{\rm mx,\bot}=&\frac{2\beta}{z}\mbox{Im}\left[\left(3-3z\partial_z+\frac{5}{4}z^2\partial_z^2-\frac{1}{4}z^3\partial^3_z\right)f\right],
\end{align}
\end{subequations}
and we have defined the function $f$
\begin{align}
f=&2\psi^{(-2)}\left(1+\frac{2iz}{\beta}\right)-\psi^{(-2)}\left[1+\frac{2iz}{\beta}\left(1-\frac{\tau}{2z}\right)\right]-\psi^{(-2)}\left[1+\frac{2iz}{\beta}\left(1+\frac{\tau}{2z}\right)\right],
\end{align}
in terms of polygamma functions $\psi^{(j)}$ \cite{delorenci2019b}. The interpretation of the functions \eqref{velmixed} is the same as the ones in Eqs.~\eqref{gene}, namely, $(\delta v_x)^2_{\rm mx,\parallel}$ apart from the factor $p^2/(16\pi^2m^2z^4)$ represents the particle velocity dispersion along the $x$ axis for a dipole oriented along the $x$ axis due to the mixed fluctuations, and similarly to $(\delta v_x)^2_{\rm mx,\bot}$, $(\delta v_y)^2_{\rm mx,\parallel}$, $(\delta v_z)^2_{\rm mx,\parallel}$, and $(\delta v_z)^2_{\rm mx,\bot}$. 

\begin{figure}[tbp]
\center
\includegraphics[width=0.5\textwidth]{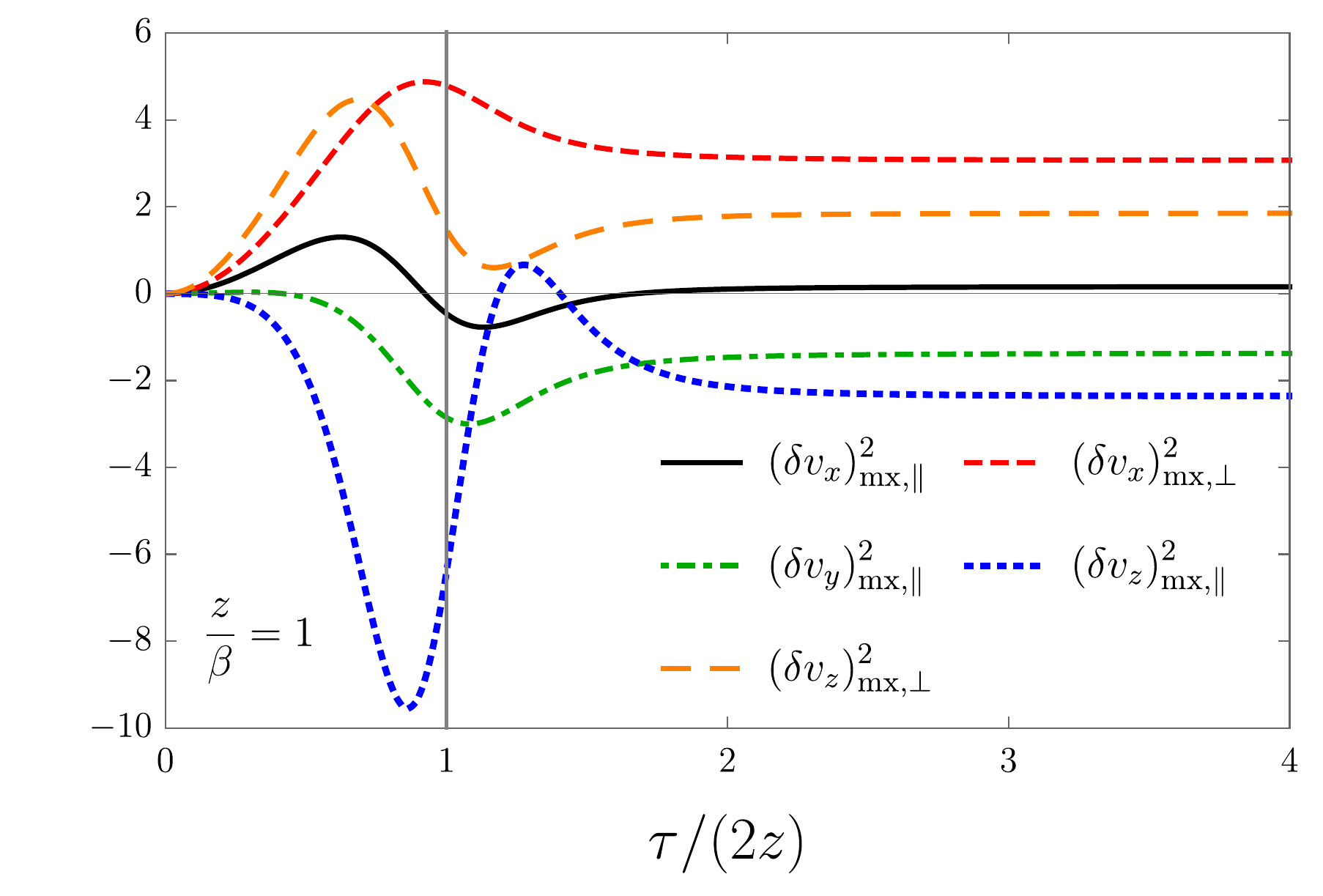}
\caption{Translational velocity dispersions (up to a factor) induced by mixed field correlations as function of $\tau/(2z)$. We set $z/\beta=1$ in these simulations. We observe that no divergence occurs at $\tau=2z$ in contrast to the $T=0$ case. Furthermore, we note that the mixed field correlations can lead to asymptotic negative velocity dispersion (dot-dashed and dotted curves).}
\label{figE}
\end{figure}

Figure \ref{figE} presents plots of the dispersions.
We note first that no divergence is observed for the mixed contributions in comparison to the zero temperature dispersions, a feature also observed for the electric monopole case \cite{delorenci2019b}. Furthermore, a clear transient regime is observed in the plots of Figure~\ref{figE} around $\tau=2z$ as the particle is placed near the wall before the stationary regime is reached. We also note that the latter can be formally calculated from our equations by taking the limit $\tau\rightarrow\infty$ in Eqs.~\eqref{velmixed}, although the final expression does not provide further information in contrast to the thermal and mirror asymptotic contributions discussed in the previous subsections.

\subsubsection{Angular velocity dispersions}

Finally, the mixed contributions of the field correlations induce angular velocity dispersions are given by 
\begin{align}
&\langle\omega_{x'}^2\rangle_{\rm mx}=\frac{p^2}{16I^2\pi^2z^2}\left[(\delta\omega)^2_{\rm mx,\parallel}\cos^2\theta+(\delta\omega)^2_{\rm mx,\bot}\sin^2\theta\right],\\
&\langle\omega_{y'}^2\rangle_{\rm mx}=\frac{p^2}{16I^2\pi^2z^2}(\delta\omega)^2_{\rm mx,\bot},
\end{align}
and
\begin{subequations}
\label{angvelmix}
\begin{align}
&(\delta\omega)^2_{\rm mx,\parallel}=\frac{2\beta}{z}\mbox{Im}\left[(1-z\partial_z)f\right],\\
&(\delta\omega)^2_{\rm mx,\bot}=\frac{\beta}{z}\mbox{Im}\left[(1-z\partial_z+z^2\partial^2_z)f\right],
\end{align}
\end{subequations}
following the same notation as in Eqs.~\eqref{velmixed}. The functions $(\delta\omega)^2_{\rm mx,\parallel}$ and $(\delta\omega)^2_{\rm mx,\bot}$ are presented in Figure~\ref{figE}.		
\begin{figure}[tbp]
\center
\includegraphics[width=0.5\textwidth]{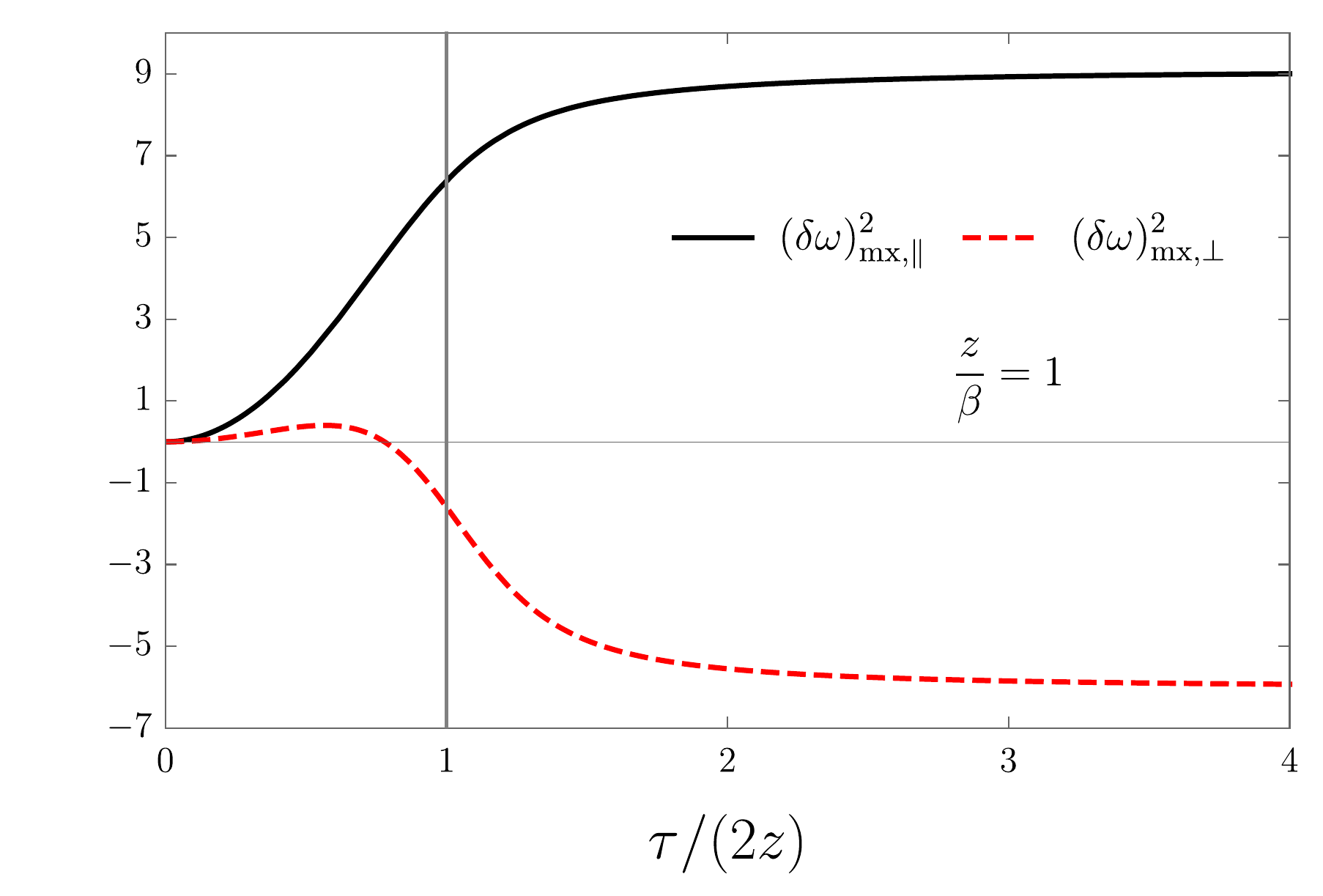}
\caption{Angular velocity dispersions (up to a fator) induced by mixed field correlations as function of $\tau/(2z)$. We set $z/\beta=1$ in these simulations. We observe that mixed correlations can also lead to negative angular velocity dispersions.}
\label{figF}
\end{figure}
We note that when $T\rightarrow0$, or equivalently $\beta\rightarrow\infty$, all the functions in Eqs.~\eqref{velmixed} and \eqref{angvelmix} vanish similarly to the thermal fluctuations discussed in subsection \ref{thermalfluc}. Furthermore, inspection of the plots in Figure~\ref{figF} (and also in Figure~\ref{figE}) reveals a remarkable feature of the dispersions when the mixed contributions are present, namely, the emergence of a regime where the dispersions become negative. In general, a subvacuum phenomenon occurs when a classically positive quantity becomes negative at a quantum level after a renormalization prescription is adopted, and the fact that the mixed dispersions can assume negative values as $\tau\gg1$ suggests that subvacuum effects might be present in this system. Nevertheless, we stress that the separation of the field correlations in a thermal, wall, and mixed contributions is artificial, and thus velocity fluctuations coming from the mixed part of the correlations are always accompanied by thermal and wall contributions.

\section{Particle energy}

\begin{figure}[tbp]
\center
\includegraphics[width=0.5\textwidth]{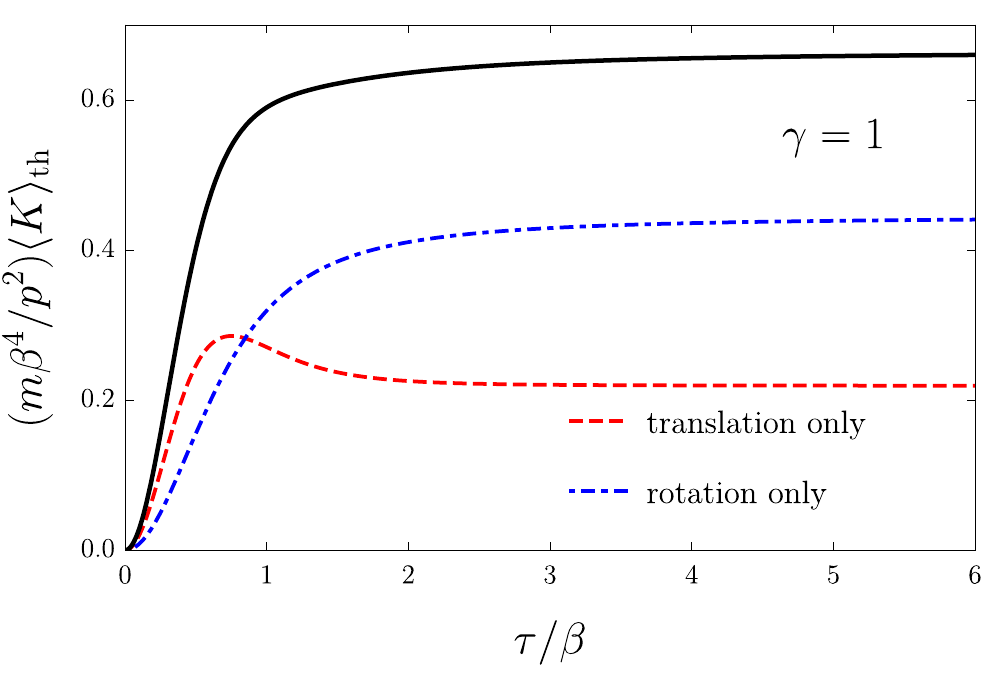}
\caption{Expectation value of the kinetic energy of the dipole in a radiation field. The behaviours of the rotational and translational contributions are separately depicted. The asymptotic regime is shortly achieved. The asymptotic value of the rotational contribution is proportional to the square of $\gamma$ parameter.}
\label{fig1}
\end{figure}

As an application of the dispersions calculated in our work, in this section we determine the energy transferred to the particle via the expectation value of the kinetic energy of the dipole $\langle K\rangle=m\langle\vv^2\rangle/2+I\langle\omega_{x'}^2+\omega_{y'}^2\rangle/2$ as function of the interaction time $\tau$, which is the sum of a translational kinetic energy plus the rotational energy defined with respect to the particle center of mass. Following the field correlations splitting into thermal, wall, and mixed contributions, we can write the kinetic energy as $\langle K\rangle=\langle K\rangle_{\rm th}+\langle K\rangle_{\rm vc}+\langle K\rangle_{\rm mx}$. Furthermore, it is instructive to discuss separately three regimes: $T\rightarrow0$, for which $\langle K\rangle=\langle K\rangle_{\rm vc}$, $z\rightarrow\infty$, which corresponds to $\langle K\rangle=\langle K\rangle_{\rm th}$, and then the combined finite $z,T$ effect. Let us consider first $\braket{K}_{\rm th}$. We find that
\begin{align}
&\braket{K}_{\rm th}=\frac{p^2}{m\beta^4}\left[\frac{\pi^2}{45}f_\beta(\tau) + \frac{4\gamma^2}{9} g_\beta(\tau)\right],
\label{energy} 
\end{align}
where $\gamma^2= m\beta^2/(4I)$.  In the above equation, the first term is identified as $\braket{m\vv^2/2}$ while the second one is $\braket{I{\boldsymbol{\omega}}^2/2}$. In the case of a dipole with a bond length $a$, its moment of inertia can be presented as $I= ma^2/4$, which leads to $\gamma = \beta/a$. As $f_\beta(\tau)$ and $g_\beta(\tau)$ have the same asymptotic value, $\gamma$ is a parameter that measures how much energy is stored by means of rotation of the dipole as compared to the energy held by its translational motion. Notice that the asymptotic value of the rotational energy is proportional to $\gamma^2$. Thus, when $\gamma \gg 1$ the translational energy is completely negligible when compared to the rotational one. Figure~\ref{fig1} depicts $\braket{K}_{\rm th}$ (solid curve) as function of $\tau/\beta$. The value $\gamma =1$ was chosen just to be possible to visualise the behaviour of the different contributions (translational and rotational) to the kinetic energy in a same figure.
It is interesting to observe that the translational contribution (dashed curve in Figure~\ref{fig1}) achieves a transient value that is larger than its asymptotic value. It means that after the dipole is placed in contact with the thermal bath of photons on average it initially gains more energy than it keeps when it achieves the stationary regime. It should be noticed that the mean values of the components of the particle velocity are related to the spatial derivatives of the electric field. Hence, the corresponding dispersions depend on the time evolution of the spatial variation of the field fluctuations, which are larger at the transition between vacuum and thermal states. Finally, part of the energy initially transferred to the linear motion of the dipole eventually returns to the radiation field or is converted in rotational energy. 

In order to have some estimates of the relative magnitude of the distinct energy contributions in Eq.~(\ref{energy}), let us express $\gamma$ as,
\begin{align}
\gamma=\frac{1}{aT}=2.29\times 10^6 \left(\frac{1{\rm nm}}{a}\right) \left(\frac{1{\rm K}}{T}\right).
\nonumber
\end{align}
As we see $\gamma \gg 1$ for most realistic configurations. At room temperatures of about $300 {\rm K}$ it follows that $\gamma \approx  10^3$ for typical molecular dipoles, which confirms that the energy absorbed by means of rotation is much greater than the energy absorbed by means of translational movement.

\begin{figure}[tbp]
\center
\includegraphics[width=0.5\textwidth]{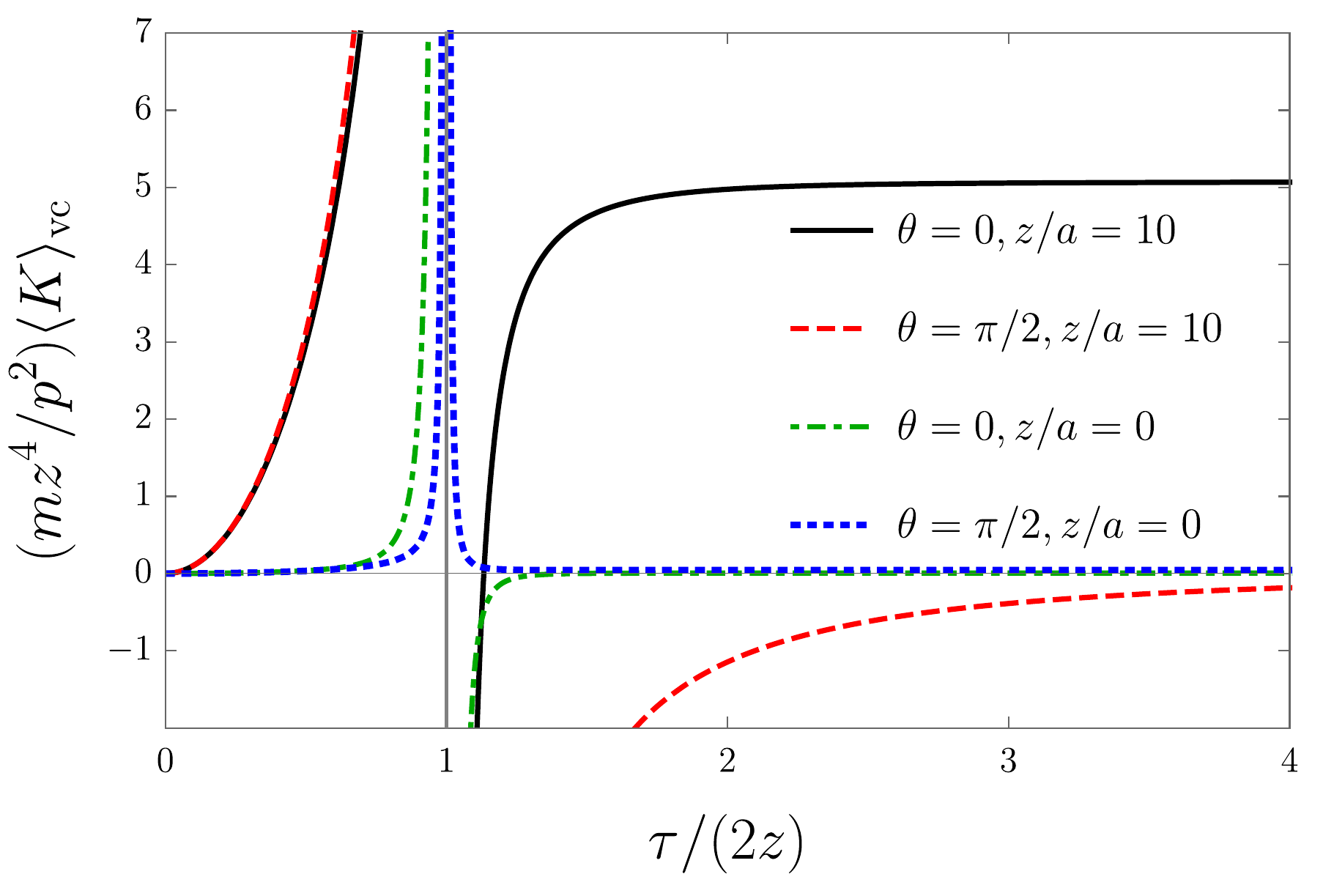}
\caption{Particle kinetic energy for $T=0$, i.e., only fluctuations coming from the presence of the wall, for two dipole orientations $\theta=0,\pi/2$. It is also here depicted the contribution of the translational kinetic energy alone ($z/a=0$), which is seen to be the subdominant contribution to the total energy, similarly to the case of pure thermal fluctuations.}
\label{figG}
\end{figure}

We now consider the zero temperature limit ($T\rightarrow0$), for which we find that the relevant particle kinetic energy is given by
\begin{align}
\braket{K}_{\rm vc}=&\frac{p^2}{16\pi^2mz^4}\bigg\{\left[(\delta \vv)^2_{\rm vc,\parallel}\cos^2\theta+(\delta \vv)^2_{\rm vc,\bot}\sin^2\theta\right]\nonumber\\
&+\frac{4z^2}{a^2}\left[(\delta \omega)^2_{\rm vc,\parallel}\cos^2\theta+(\delta \omega)^2_{\rm vc,\bot}(1+\sin^2\theta)\right]\bigg\},
\label{energyv}
\end{align}
where we set $(\delta \vv)^2_{\rm vc,\parallel}=(\delta v_x)^2_{\rm vc,\parallel}+(\delta v_y)^2_{\rm vc,\parallel}+(\delta v_z)^2_{\rm vc,\parallel}$, $(\delta \vv)^2_{\rm vc,\perp}=2(\delta v_x)^2_{\rm vc,\perp}+(\delta v_z)^2_{\rm vc,\perp}$, and, again, $a$ denotes the dipole bond length. We observe that similarly to what occurs for the thermal contribution alone through the parameter $\gamma$, the rotational kinetic energy when compared to the translational energy is multiplied by the factor $z^2/a^2$. For the sake of illustration, if we set $z/a=10$, Figure~\ref{figG} shows that the rotational energy is dominant throughout the system evolution. 
Moreover, we notice that both $\langle K\rangle_{\rm th}$ and $\langle K\rangle_{\rm vc}$ tend asymptotically to non-negative constants as $\tau\rightarrow\infty$. Indeed, we obtain from Eqs.~\eqref{energy} and \eqref{energyv} that
\begin{align}
&\langle K\rangle_{\rm th}\rightarrow\frac{p^2}{m\beta^4}\left(\frac{\pi^2}{45}+\frac{4\gamma^2}{9}\right),\label{enlate1}\\
&\langle K\rangle_{\rm vc}\rightarrow\frac{p^2}{16\pi^2mz^4}\left[4-3\cos2\theta+\frac{4z^2}{a^2}(1+\cos2\theta)\right],\label{enlate2}
\end{align}
as $\tau\rightarrow\infty$. In particular, no subvacuum effect is observed after the energy exchange from the quantum vacuum to the particle occurs, following the same conclusion observed for the electric monopole \cite{ford2004}.

\begin{figure}[tbp]
\center
\includegraphics[width=0.5\textwidth]{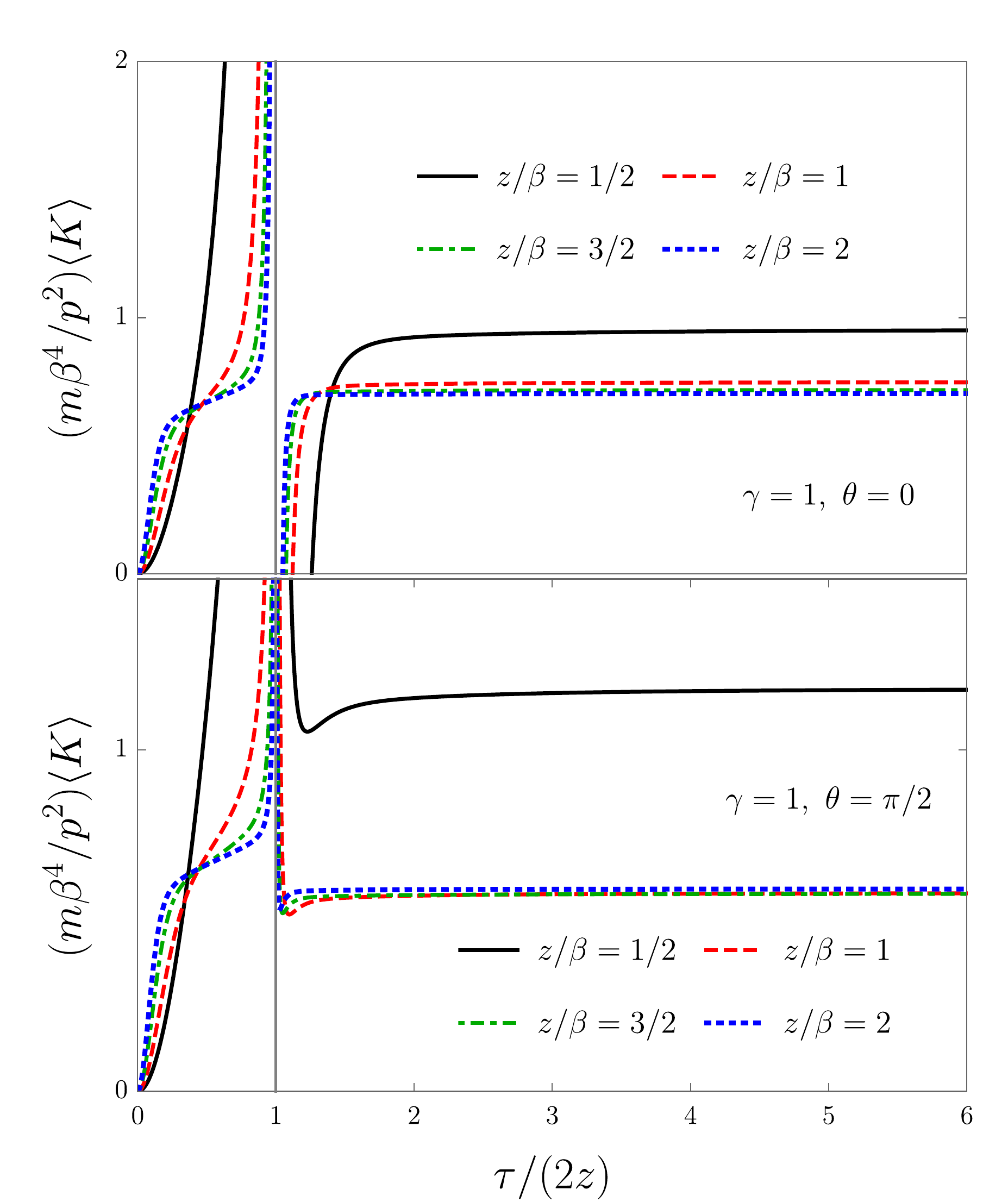}
\caption{Particle kinetic energy gained by the work done by the quantum fluctuations as function of $\tau/(2z)$ for several values of $z/\beta$. Upper panel: Energy of the particle whose dipole moment is oriented parallel to the wall. Lower panel: Particle energy for the dipole moment oriented perpendicular to the wall. We observe that for both orientations immediately after $\tau=2z$ the particle receives more energy on average than it holds asymptotically.
}
\label{figH}
\end{figure}

When the particle is in the presence of both the wall and the thermal bath its energy profile is richer. It is instructive to work with the dimensionless quantity $(m\beta^4/p^2)\langle K \rangle$, which can be straightforwardly calculated from the dispersions discussed in the previous sections. Also, the dependence of $(m\beta^4/p^2)\langle K \rangle$ on the parameters $z,\tau$, and $\beta$ occurs through $z/\beta$ and $\tau/(2z)$, and typical values of $z/\beta$ can be specified by returning to dimensionful units
\begin{equation}
zT=4.37 \times10^{2}\left(\frac{z}{1\mbox{m}}\right)\left(\frac{T}{1\mbox{K}}\right),
\end{equation}
implying that $z/\beta$ is of order $1$ for $z\sim1\ \mbox{mm}$ and $T\sim10\ \mbox{K}$. We show in Figure~\ref{figH} our findings for the energy gained by the particle.
Figure \ref{figH} upper panel depicts the particle energy as function of $\tau/(2z)$ for several values of $z/\beta$ for a dipole aligned parallel to the wall ($\theta=0$). Salient features include the positive asymptotic value reached by the energy and the quick stabilisation after $\tau=2z$.
 Similar behaviour is observed for the dipole oriented perpendicularly ($\theta=\pi/2$) to the conducting plane (Figure~\ref{figH} lower panel).

 Furthermore, no residual subvacuum effects, that would correspond to negative residual kinetic energies in our case, are observed in Figure~\ref{figH}. Such an effect was reported in \cite{delorenci2019b} for the electric monopole, and it is interpreted as a quantum cooling effect, where the work done by the vacuum fluctuations over the particle diminishes its overall positive energy.
Notwithstanding, a subtle quantum cooling effect can be observed in the particle residual energy ($\tau\rightarrow\infty$) when its dipole moment is oriented perpendicularly to the wall. Indeed, we note that if the particle is placed either in the thermal bath or near the wall, its residual energy is always positive, i.e., $\langle K\rangle_{\rm th},\langle K\rangle_{\rm vc}>0$ [cf. Eqs.~\eqref{enlate1} and \eqref{enlate2}]. 
\begin{figure}[tbp]
\center
\includegraphics[width=0.5\textwidth]{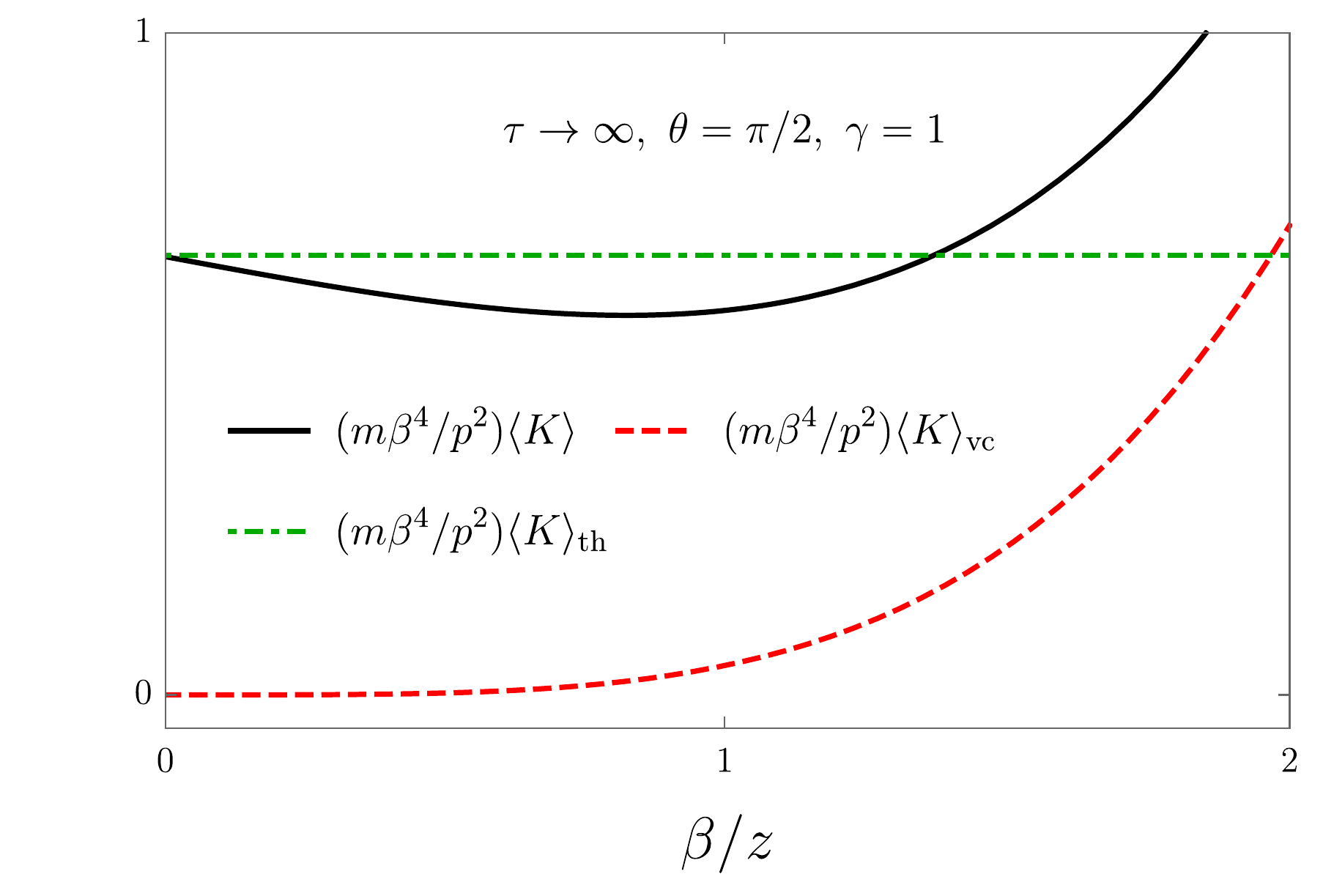}
\caption{Comparison between the residual kinetic energies $\langle K\rangle_{\rm th}$, $\langle K\rangle_{\rm vc}$, and $\langle K\rangle$ as functions of $\beta/z$. We observe that when both the thermal bath and the wall are present, $\langle K\rangle<\langle K\rangle_{\rm th}$ for a range of $\beta/z$. This shows that subvacuum effects are present, characterised by an energy content being extracted from the particle by the wall, and corresponding to a quantum cooling effect.}
\label{figI}
\end{figure}
However, we find that when the particle is in the presence of both the wall and the thermal bath, $\langle K\rangle<\langle K\rangle_{\rm th}$ for $\beta/z\sim1$ as shown in Figure~\ref{figI}, indicating that a conducting wall can lead to a decrease in the kinetic energy of an electric dipole in a thermal bath, depending on the particle's distance to the wall.

\section{Estimates}
\label{estimates}

In order to obtain estimates and address the validity of our assumptions in deriving Eqs.~\eqref{vi}, \eqref{velang1}, \eqref{velang2}, let us consider the case of pure thermal bath, and take the particle to have mass equal to $m_{{}_{\text{o}}}~=~1.2\times 10^{-22}{\rm g}$ and a dipole moment $p_{{}_{\text{o}}}=10.27{\text D}$, where ${\text D}$ denotes debye ($1{\text D} \approx 3.36\times10^{-30}{\rm C m}$). These are approximately the data for typical molecules like the potassium chloride (KCl), which are used here only for the sake of comparison.

After the system achieves the asymptotic regime, that occurs in a temperature dependent time interval of about 
\begin{equation}
\tau_{\rm e}=10^{-11}\left(\frac{\text{1K}}{T}\right) {\rm s}\label{crit}
\end{equation}
after it is placed in contact with the thermal environment, the uncertainty in its linear $\Delta v$ and angular $\Delta \omega$ velocities can be obtained directly from the square-roots of Eqs. (\ref{v}) and (\ref{omega}), respectively. For the linear velocity we find
\begin{align}
\Delta v = 7.2\times 10^{-15} \left(\frac{p}{{ p_{\text o}}}\right) \left(\frac{{ m_{\text o}}}{m}\right)\left(\frac{T}{1{\rm K}}\right)^2{\rm ms^{ -1}},
\nonumber
\end{align}
which, at room temperature $T = 300 {\rm K}$, is approximately $10^{-9} {\rm ms^{-1}}$. On the other hand the uncertainty in the angular velocity is
\begin{align}
\Delta \omega = 4.7\times 10^{3} \left(\frac{p}{{ p_{\text o}}}\right) \left(\frac{{ m_{\text o}}}{m}\right)\left(\frac{T}{1{\rm K}}\right)\left(\frac{1\text{\AA}}{a}\right)^2{\rm rad\, s^{-1}},\label{angdis}
\end{align}
which achieves a value of the order of $10^{5}\, {\rm rad}\, {\rm s^{-1}}$  at room temperature.

A possible observable in this system is the radiation emitted by the dipole rotation. If the dipole is initially placed at rest in the bulk of a cavity with a fixed temperature $T$, it is expected that after the system reaches its stationary regime, the uncertainty in its rotation frequency will be $\Delta\omega$, whose estimate can be obtained from the formula above. Therefore, it is expected that radiation may be emitted within a frequency interval $0 \le \omega \le \Delta\omega$. The averaged power $P$ radiated by the system can be estimated by assuming an idealised model of an electric dipole rotating in a plane with angular velocity $\omega$. As shown in standard textbooks \cite{landau}, the dominant contribution to this quantity is given by $P = p^2\omega^4/(6\varepsilon_0 c^3)$. Hence, the mean value of the corresponding quantum observable can be estimated as $\langle P\rangle \sim 10^{-46 }{\rm Js^{-1}}$ for system above examined in STP conditions,
which is a tiny effect as expected. It should be stressed, however, that this {\it is} the radiation emitted by a single dipole due to quantum fluctuations of the electromagnetic field.

Another possible way of measuring the vacuum fluctuations here discussed is based on a different system preparation in which a molecule with a dipole moment in a given initial direction traverses the region containing the thermal bath near a flat wall. As quantum dispersions of the velocity are not equal in all directions the dipole will present an anisotropic dispersion in its direction of motion. In this experiment, the deviation with respect to the classical direction of motion can be detected and compared with the velocity uncertainty acquired during its interaction with the quantum fluctuations. Despite the fact that this procedure requires a moving particle, the expressions derived here for the dispersions remain valid as long as the classical motion is nonrelativistic. This reasoning was used as a possible way of detecting modified vacuum fluctuations in a similar system \cite{delorenci2016}. 

We conclude this section with a discussion concerning the limit of validity of our results. We recall that in deriving equations \eqref{vi}, \eqref{velang1}, and \eqref{velang2} we assumed that both the particle's position and its dipole moment remain time independent throughout the system evolution. However, in Eq.~\eqref{angdis} for instance we showed that for parameters within experimental range in $1$ second we expect the dipole moment to undergo many rotations, thus violating the model's assumption of constant $\p$. In particular, our results cannot be extrapolated to study (thermodynamic) equilibrium conditions in contrast to the classical Brownian motion analysis \cite{li2010}. Therefore, our results are reliable for a certain measuring time $\tau$ determined by the system parameters. From our estimates for a pure thermal bath, we obtained for instance that the system transient time is given by Eq.~\eqref{crit}, from which we can estimate the angle of rotation using the angular velocity on Eq.~\eqref{angdis} as $\tau_{\rm e}\Delta \omega  \sim 10^{-8} $ rad, a certainly small quantity that justifies treating $\p$ as constant in Eqs.~\eqref{vi}, \eqref{angx}, and \eqref{angy}. Finally, we also assumed that the particle's center of mass remains approximately at rest during the system evolution. In \cite{delorenci2016} it was shown that as long as the particle undergoes nonrelativistic motion, which in our model corresponds to small velocity dispersions, assuming a stationary center of mass is equivalent to determining the leading contribution to the dispersions.

\section{Final remarks}

At the stochastic level, the fluctuations of an isotropic gas of photons induce an anisotropic motion of a dipole, the fluctuations being greater in the direction perpendicular to the orientation of the dipole moment, unlike the case of a charged particle, where the motion induced by a gas of photons is isotropic \cite{delorenci2019b,yu2006,jt2009}.
This is an interesting result, because even though the dipole introduces a preferential direction, the driven stochastic force comes from an isotropic system. This effect is similar to the case of a charged particle initially at rest near a perfectly reflecting wall. Vacuum fluctuations of the modified vacuum state will produce an uncertainty in the parallel component of the particle velocity \cite{ford2004,delorenci2016}. Furthermore, as in the case of the charged particle, the thermal fluctuations on the dipole dispersion are positive, regardless of the system temperature.

When dispersive effects are neglected, fluctuations of thermodynamic quantities usually have a random walk behaviour being proportional to the interaction time \cite{ford2005}. In such cases, the thermal reservoir continuously gives away energy to the system, increasing its motion, and a dissipative force is needed so that this energy is given back to the environment and the dispersions settle to their usual thermal equilibrium value where the state of the particle is also a Gibbs state, with the same temperature as the environment. 
Nonetheless, in this model the thermal environment only gives away a finite amount of energy to the dipole. In Ref.~\cite{camargo2021} it was shown that this is so because the fluctuations of the gas of photons have anti-correlations, which, when integrated over an infinite interaction time, amount to a finite positive contribution. These anti-correlations arise because, contrarily to the usual thermal case, the field does not only push, but also pulls the dipole. Moreover, it dismisses the discomfort due to the lost energy when the fluctuations in the dipole velocity decay from the peak to a constant late-time value.

Furthermore, due to the coupling with the field, which is not negligible when compared to the free dipole Hamiltonian, the late-time equilibrium state of the dipole is not a Gibbs thermal state \cite{hu2018}. The effective temperature, given by the velocity fluctuations, is different from the temperature of the KMS state of the electric field. This highlights the difference with a usual statistical mechanical system.

\appendix
\section{Field correlations and the relevant formulas}
\label{appA}

The field correlations for the electromagnetic vacuum state under consideration can be found, for instance, in reference \cite{Brown}. At finite temperature and in the presence of the conducting wall at $z=0$ we find that the renormalized correlations split as $\langle E_{x}(t,\x)E_{x}(t',\x')\rangle=\langle E_{x}(t,\x)E_{x}(t',\x')\rangle_{\rm th}+\langle E_{x}(t,\x)E_{x}(t',\x')\rangle_{\rm vc}+\langle E_{x}(t,\x)E_{x}(t',\x')\rangle_{\rm mx}$, and similarly for $y$ and $z$ correlations, where
\begin{align}
\langle E_{x}(t,\x)E_{x}(t',\x')\rangle_{\rm vc}=&-\frac{1}{\pi^2}\frac{(\Delta t)^2-(\Delta x)^2+(\Delta y)^2+(\hat{\Delta} z)^2}{[(\Delta t)^2-(\Delta x)^2-(\Delta y)^2-(\hat{\Delta} z)^2]^3},\nonumber\\
\langle E_{x}(t,\x)E_{x}(t',\x')\rangle_{\rm th}=&\frac{2}{\pi^2}\mbox{Re}\left[\sum_{n=1}^{\infty}\frac{(\Delta t-in\beta)^2-(\Delta x)^2+(\Delta y)^2+(\Delta z)^2}{[(\Delta t-in\beta)^2-(\Delta x)^2-(\Delta y)^2-(\Delta z)^2]^3}\right],\nonumber\\
\langle E_{x}(t,\x)E_{x}(t',\x')\rangle_{\rm mx}=&-\frac{2}{\pi^2}\mbox{Re}\left[\sum_{n=1}^{\infty}\frac{(\Delta t-in\beta)^2-(\Delta x)^2+(\Delta y)^2+(\hat{\Delta} z)^2}{[(\Delta t-in\beta)^2-(\Delta x)^2-(\Delta y)^2-(\hat{\Delta} z)^2]^3}\right],\nonumber\\
\langle E_{z}(t,\x)E_{z}(t',\x')\rangle_{\rm vc}=&\frac{1}{\pi^2}\frac{(\Delta t)^2+(\Delta x)^2+(\Delta y)^2-(\hat{\Delta} z)^2}{[(\Delta t)^2-(\Delta x)^2-(\Delta y)^2-(\hat{\Delta} z)^2]^3},\nonumber\\
\langle E_{z}(t,\x)E_{z}(t',\x')\rangle_{\rm th}=&\frac{2}{\pi^2}\mbox{Re}\left[\sum_{n=1}^{\infty}\frac{(\Delta t-in\beta)^2+(\Delta x)^2+(\Delta y)^2-(\Delta z)^2}{[(\Delta t-in\beta)^2-(\Delta x)^2-(\Delta y)^2-(\Delta z)^2]^3}\right],\nonumber\\
\langle E_{z}(t,\x)E_{z}(t',\x')\rangle_{\rm mx}=&\frac{2}{\pi^2}\mbox{Re}\left[\sum_{n=1}^{\infty}\frac{(\Delta t-in\beta)^2+(\Delta x)^2+(\Delta y)^2-(\hat{\Delta} z)^2}{[(\Delta t-in\beta)^2-(\Delta x)^2-(\Delta y)^2-(\hat{\Delta} z)^2]^3}\right],\nonumber
\end{align}  
$\Delta a:=a-a'$, $\hat{\Delta} a:=a+a'$, the $y$ correlations are obtained from the $x$ correlations by exchanging $\Delta x\leftrightarrow\Delta y$, and $\Delta t=t-t'+i\epsilon$, with $\epsilon>0$ to be taken to zero at the end of the calculation. The crossed components of the correlation functions were not listed above because they will all vanish in the limit of point coincidence, including their derivatives that appear in Eqs.~(\ref{corv}), (\ref{corw1}) and (\ref{corw2}). All the dispersions calculated in our work can be found with the aid of the following two integrals
\begin{align}
\int_0^\tau dt\int_0^\tau dt'\frac{(\Delta t)^2-b^2}{[(\Delta t)^2-c^2]^3}=\frac{\tau^2(b^2-c^2)}{4c^4(c^2-\tau^2)}-\frac{c^2+3b^2}{16c^5}\tau\ln\left(\frac{c-\tau}{c+\tau}\right)^2,
\end{align}   
and
\begin{align}
\sum_{n=1}^{\infty}\int_0^\tau dt\int_0^\tau dt'\frac{(\Delta t-in\beta)^2-b^2}{[(\Delta t-in\beta)^2-c^2]^3}=&-\frac{1}{8\beta^2}\left(1+\frac{c^2-b^2}{4c}\frac{\partial\ }{\partial c}\right)\bigg[s\left(1+\frac{i\tau}{\beta},\frac{ic}{\beta}\right)\nonumber\\
&+s\left(1-\frac{i\tau}{\beta},\frac{ic}{\beta}\right)-2s\left(1,\frac{ic}{\beta}\right)+c\leftrightarrow -c\bigg],
\end{align}   
with the function $s(a,b)$ is defined in terms of polygamma functions $\psi^{(j)}$ as
\begin{align}
s(a,b)=-\frac{1}{b}\psi^{(0)}(a)+\frac{2}{b^2}\psi^{(-1)}(a+b)-\frac{2}{b^3}[\psi^{(-2)}(a+b)-\psi^{(-2)}(a)].
\end{align}
The simplifications used to obtain the equations involving polygamma functions in the main text explore the defining identity $d\psi^{(j)}(x)/dx=\psi^{(j+1)}(x)$.

\acknowledgments

This work was partially supported by the Brazilian research agencies CNPq (Conselho Nacional de Desenvolvimento Cient\'{\i}fico e Tecnol\'ogico) under Grant No. 305272/2019-5, CAPES (Coordena\c{c}\~ao de Aperfei\c{c}oamento de Pessoal de N\'{\i}vel Superior), and FAPES (Funda\c{c}\~ao de Amparo a Pesquisa do Esp\'{\i}rito Santo).

\end{document}